\documentclass[fleqn,showpacs,aps,graphicx]{revtex4}
\usepackage{graphicx}
\usepackage{mathrsfs}
\usepackage{amsfonts}
\usepackage{amssymb}
\usepackage[mathscr]{eucal}
\usepackage{amssymb}
\usepackage{amsmath}
\usepackage{dcolumn}
\usepackage{bm}

\begin{document}

\title{Efficient entanglement concentration  for arbitrary less-entangled NOON state assisted with single photon}

\author{Lan Zhou$^{1}$, Yu-Bo Sheng,$^{2,3}$}
\address{$^1$ College of Mathematics \& Physics, Nanjing University of Posts and Telecommunications, Nanjing,
210003, China\\
$^2$ Institute of Signal Processing  Transmission, Nanjing
University of Posts and Telecommunications, Nanjing, 210003,  China\\
$^3$Key Lab of Broadband Wireless Communication and Sensor Network
 Technology, Nanjing University of Posts and Telecommunications, Ministry of
 Education, Nanjing, 210003, China\\}

\date{\today }

\begin{abstract}

We put forward two efficient entanglement concentration protocols (ECPs) for distilling the maximally entangled NOON state from arbitrary less-entangled NOON state with only an auxiliary single photon. With the help of the weak cross-Kerr nonlinearities, both the two ECPs can be used repeatedly to get a high success probability. In the first ECP, the auxiliary single photon should be shared by the two parties say Alice and Bob. In the second ECP, the auxiliary single  photon is only possessed by Bob, which can greatly increase the practical success probability by avoiding the transmission loss. Moreover, Bob can operate the whole protocol alone, which makes the protocol more simple.  Therefore, our two ECPs, especially the second ECP may be more useful and convenient in the current quantum information processing.
\end{abstract}
\pacs{ 03.67.Dd, 03.67.Hk, 03.65.Ud} \maketitle

\section{Introduction}

  Entanglement plays a significant role in the quantum information mechanics, for it not only can hold the power
for the quantum nonlocality, but
also can provide wide applications in the quantum information processing (QIP) \cite{book,rmp}. For example, in the fields of quantum dense coding \cite{code}, quantum teleportation \cite{teleportation}, quantum key distribution \cite{Ekert91,QKDdeng1,QKDdeng2} and other quantum information processes \cite{QSDC1,QSDC2,QSDC3,QSS1,QSS2,QSS3,communication,communication1}, we need entangled quantum systems to setup the
quantum channel. Recently, a special quantum state, which is so-called the NOON state has attracted great attention in the field of quantum information \cite{noon,noon1,noon2,noon3,noon4,noon5,noon6,noon7}. The NOON state can be written with the form
\begin{eqnarray}
|NOON\rangle_{ab}=\frac{1}{\sqrt{2}}(|N,0\rangle_{ab}+|0,N\rangle_{ab}),\label{NOON}
\end{eqnarray}
which describes the state where all $N$ particles are in the spatial mode $a$ and none is in the spatial mode $b$, plus none in the mode $a$ and all in the mode $b$. Due to its unique properties, the NOON state has provided many important applications. For example, it has been proved that the NOON state can serve as an important resource for Heisenberg-limited metrology and quantum lithography, where it is greatly sensitivity for optical interferometry and can approach the Heisenberg limit of 1/N \cite{metrology,metrology1,metrology2}. Meanwhile, the NOON state also shows
the de Broglie wavelength of $\lambda/N$ for $N$-photon interference \cite{noon5,wavelength1,wavelength2,wavelength3}. It is
worth noting that the power of NOON state lies in its entanglement. In all applications, the ideal NOON state should be the maximally entangled NOON state as Eq. (\ref{NOON}). However, in practical application process, since the decoherence is unavoidable during the storage and transmission of particles over noisy channels, the quality
of entanglement can be easily degraded. The noise may lead the maximally entangled NOON state degrade to the less-entangled state. The general form of the less-entangled NOON state can be written as
\begin{eqnarray}
|NOON'\rangle_{ab}=\alpha|N,0\rangle_{ab}+\beta|0,N\rangle_{ab},\label{NOON1}
\end{eqnarray}
 where $\alpha$ and $\beta$ are the entanglement coefficients, $|\alpha|^{2}+|\beta|^{2}=1$, and $\alpha\neq\beta$. In practical application, if the maximally entangled NOON state is polluted, we need to recover it into the maximally entangled NOON state.

 The method for distilling the less-entangled pure state into the maximally entangled state is called the entanglement concentration \cite{C.H.Bennett2}. In 1996, Bennett \emph{et al.} proposed the first entanglement concentration protocol (ECP), which is known as the Schmidt projection method \cite{C.H.Bennett2}. This protocol is regarded to be an exciting start for the study of the entanglement concentration. Since then, various interesting ECPs have been put forward, successively \cite{swapping1,swapping2,zhao1,Yamamoto1,shengpra2,shengqic,shengpra3,shengwstateconcentration,shengpla,bose,wangxb,wangc2,dengpra,wanghf2,yeliu,gubin,dengsingle}. For example, In 1999, Bose \emph{et al.} proposed an ECP based on entanglement swapping \cite{swapping1}, which was later improved by Shi \emph{et al.} \cite{swapping2}. In 2001, Zhao et al. and Yamamoto et al. proposed two similar concentration protocols based on polarizing beam splitters (PBSs) independently \cite{zhao1,Yamamoto1}. In 2010, the ECP for single-photon entanglement was proposed by Sheng \emph{et al.} \cite{shengqic}. Recently, the ECPs for less-entangled W states were proposed \cite{shengwstateconcentration,wanghf2,gubin,dengsingle}. So far, most of the ECPs
are focused on the two-particle entanglement, especially the encoding system in the polarization degree of freedom in the optics system.
They cannot deal with the less-entangled NOON state. Recently, we have proposed an ECP for less-entangled NOON state \cite{zhounoon}. However, this protocol is not optimal. In the protocol, two pairs of less-entangled NOON states are required, and after performing this ECP, at least one pair of maximally entangled NOON state can be obtained.

In the paper, we put forward two efficient ECPs for concentrating the less-entangled NOON state into the maximally entangled NOON state. Different from the
 ECP described in Ref. \cite{zhounoon}, our two ECPs only require one pair of less-entangled NOON state and a single auxiliary photon to complete the concentration task and can reach the same success probability as Ref. \cite{zhounoon}. In the two ECPs, we adopt the weak cross-Kerr nonlinearity to construct the quantum nondemolition (QND) measurement, which makes both the two ECPs can be realized in current technology.
 Moreover, we will prove that both the two ECPs can be used repeatedly to get a high success probability.
 The first ECP is essentially inspired by the ECP in Ref. \cite{shengqic}, as the single-photon entanglement is the simplest NOON state.  The second ECP is more optimal than the first one, for it only requires the
local single auxiliary photon, which can greatly avoid the photon loss during the transmission of the single auxiliary photon. Moreover, only Bob needs to perform the whole task which can simplify the practical experiment largely.

  This paper is organized as follows: In Sec. II and Sec. III, we explain the basic principles of the first and the second ECP, respectively. In Sec. IV, we make a discussion and summary.

\section{The first concentration protocol for the N-photon NOON state}
The cross-Kerr nonlinearity is the key element for both two ECPs. Therefore, before describing the first ECP, we would like to make a brief introduction on the cross-Kerr nonlinearity. The cross-Kerr nonlinearity has been widely used in the field of quantum information. It has played an important role in the construction of CNOT gate \cite{gate,gate1}, Bell-state analysis \cite{bell,bell2}, and so on \cite{lin2,lin3,he1,he2,he3}. Especially, the cross-Kerr nonlinearity is a powerful tool in the construction of the QND, which is widely used in the quantum entanglement concentration \cite{shengwstateconcentration,shengpra2,shengpra3,shengqic,dengpra,wanghf2,yeliu,gubin,dengsingle}. The cross-Kerr nonlinearity can be described by its Hamiltonian as
\begin{eqnarray}
H_{ck} = \hbar\chi\hat{n_{a}}\hat{n_{b}},
\end{eqnarray}
where $\hbar\chi$ is the coupling strength of the nonlinearity, which depends on the cross-Kerr material \cite{cross1,gate,gate1}. $\hat{n_{a}}$ and $\hat{n_{b}}$ are the photon number operators for the spatial mode a and b. During the cross-Kerr interaction, a laser pulse in the coherent state $|\alpha\rangle$ interacts with photons through a proper cross-Kerr material. If the single photon is presented, it can induce a phase shift $\theta$ to the coherent state . The cross-Kerr interaction can be described as
\begin{eqnarray}
U_{ck}|\psi\rangle|\alpha\rangle&=&(\gamma|0\rangle+\delta|1\rangle)|\alpha\rangle\rightarrow\gamma|0\rangle|\alpha\rangle+\delta|1\rangle|\alpha e^{i\theta}\rangle.\label{QND}
\end{eqnarray}
Here, $|0\rangle$ and $|1\rangle$ mean none photon and one photon, respectively. $\theta$ = $\chi t$, and t means the interaction
time for the signal with the nonlinear material. According to the Eq. (\ref{QND}), it can be easily found that the phase shift of the coherent state is
directly proportional to the number of photons. Therefore, by measuring the phase shift of the coherent state, we can check the photon number without destroying the photons.
\begin{figure}[!h]
\begin{center}
\includegraphics[width=8cm,angle=0]{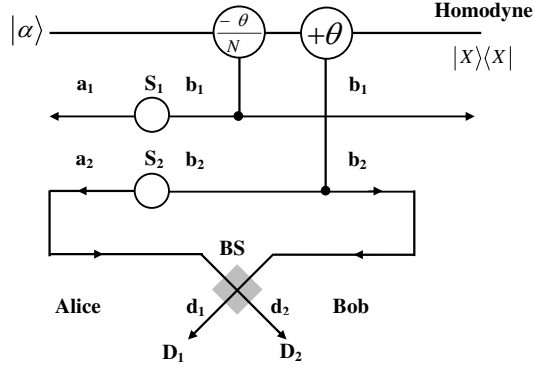}
\caption{A schematic diagram of our first ECP. In this protocol, an auxiliary photon which is shared by the two parties is used to concentrate the less-entangled  NOON state. Bob makes the photons in the spatial mode b$_{1}$ and b$_{2}$ pass through two cross-Kerr nonlinearities, respectively, and selects the items $|N,0,0,1\rangle_{a_{1}b_{1}a_{2}b_{2}}$ and $|0,N,1,0\rangle_{a_{1}b_{1}a_{2}b_{2}}$, which make the coherent state pick up $\pm\theta$. Then the two parties make the photons in the modes a$_{2}$ and b$_{2}$ pass through the 50:50 BS, and finally can obtain the maximally entangled NOON state. By making the discarded items pass through the BS, they can evolve to a new less-entangled NOON state and can be reconcentrated for the next round.}
\end{center}
\end{figure}

Now, we start to explain our first ECP for distilling the maximally entangled NOON state from arbitrary less-entangled NOON state. The principle of the ECP is shown in Fig. 1. We suppose that Alice (A) and Bob (B) share a less-entangled N-photon NOON state $|\phi\rangle_{a_{1}b_{1}}$ in the spatial modes a$_{1}$ and b$_{1}$, which is generated by the photon source S$_{1}$. $|\phi\rangle_{a_{1}b_{1}}$ can be written of the form
\begin{eqnarray}
|\phi\rangle_{a_{1}b_{1}}&=&\alpha|N,0\rangle_{a_{1}b_{1}}+\beta|0,N\rangle_{a_{1}b_{1}}.\label{oringin}
\end{eqnarray}
Here $\alpha$ and $\beta$ are the entanglement coefficients of the NOON state, where $|\alpha|^{2}+|\beta|^{2}=1$, and $\alpha\neq\beta$.

A single-photon entanglement source S$_{2}$ emits an auxiliary photon and sends it to Alice and Bob in the spatial modes a$_{2}$ and b$_{2}$, which creates a single-photon entangled state of the form
\begin{eqnarray}
|\phi\rangle_{a_{2}b_{2}}&=&\alpha|1,0\rangle_{a_{2}b_{2}}+\beta|0,1\rangle_{a_{2}b_{2}}.
\end{eqnarray}
Here, for ensuring the coefficients of the auxiliary photon, we need to know the initial coefficients of the less-entangled NOON state,  $\alpha$ and $\beta$. It is not difficult to know the values of the two coefficients by measuring
an enough amount of the sample entangled NOON state \cite{dengsingle}.

Then the whole N+1 photons system can be described as
\begin{eqnarray}
|\Phi\rangle_{a_{1}b_{1}a_{2}b_{2}}&=& |\phi\rangle_{a_{1}b_{1}}\otimes |\phi\rangle_{a_{2}b_{2}}\nonumber\\
&=& (\alpha^{2}|N,0,1,0\rangle+\alpha\beta|N,0,0,1\rangle+ \alpha\beta|0,N,1,0\rangle+\beta^{2}|0,N,0,1\rangle)_{a_{1}b_{1}a_{2}b_{2}}.\label{whole}
\end{eqnarray}

According to Fig. 1, Bob makes the photons in the spatial modes b$_{1}$ and b$_{2}$ pass through two cross-Kerr nonlinearities, respectively. Then, the N+1 photons system combined with the coherent state can evolve to
\begin{eqnarray}
|\Phi\rangle_{a_{1}b_{1}a_{2}b_{2}}\otimes|\alpha\rangle&\rightarrow&\alpha^{2}|N,0,1,0\rangle_{a_{1}b_{1}a_{2}b_{2}}|\alpha\rangle+\alpha\beta|N,0,0,1\rangle_{a_{1}b_{1}a_{2}b_{2}}|\alpha e^{i\theta}\rangle\nonumber\\
&+& \alpha\beta|0,N,1,0\rangle_{a_{1}b_{1}a_{2}b_{2}}|\alpha e^{-i\theta}\rangle+\beta^{2}|0,N,0,1\rangle_{a_{1}b_{1}a_{2}b_{2}}|\alpha\rangle.\label{QND1}
\end{eqnarray}
From Eq. ({\ref{QND1}}), we can easily find that the items $|N,0,0,1\rangle_{a_{1}b_{1}a_{2}b_{2}}$ and $|0,N,1,0\rangle_{a_{1}b_{1}a_{2}b_{2}}$ make  the coherent state pick up the phase shifts of $\theta$ and $-\theta$, respectively, while other two items make the coherent state pick up no phase shift. As $\pm \theta$ is undistinguishable during the homodyne measurement, Bob selects the items corresponding to the phase shift of $\pm\theta$, and discards the other items. Then the Eq. (\ref{whole}) can collapse to
\begin{eqnarray}
|\Phi_{1}\rangle_{a_{1}b_{1}a_{2}b_{2}}&=&\frac{1}{\sqrt{2}}(|N,0,0,1\rangle_{a_{1}b_{1}a_{2}b_{2}}+|0,N,1,0\rangle_{a_{1}b_{1}a_{2}b_{2}}),\label{pick}
\end{eqnarray}
with the success probability of $2|\alpha\beta|^{2}$.

Then Alice and Bob make the photons in the modes a$_{2}$ and b$_{2}$ pass through the 50:50 BS and take the Bell measurement, which makes
\begin{eqnarray}
\hat{a}_{2}^{\dagger}|0\rangle=\frac{1}{\sqrt{2}}(\hat{d}_{1}^{\dagger}|0\rangle-\hat{d}_{2}^{\dagger}|0\rangle)\nonumber\\ \hat{b}_{2}^{\dagger}|0\rangle=\frac{1}{\sqrt{2}}(\hat{d}_{1}^{\dagger}|0\rangle+\hat{d}_{2}^{\dagger}|0\rangle).
\end{eqnarray}
Here, a$_{j}$, b$_{j}$ and d$_{j}$ ($j=1,2$) are the creation operators for the spatial mode a$_{j}$, b$_{j}$ and d$_{j}$, respectively. The creation operators obey the rules that $\hat{i}^{\dagger}_{j}|0\rangle=|1\rangle_{i_{j}}$ and $(\hat{i}_{j}^{\dagger})^{N}|0\rangle=\sqrt{N}|N\rangle_{i_{j}}$, where $i=a,b,d$ and $j=1,2$. After the BS, Eq. (\ref{pick}) can ultimately evolve to
\begin{eqnarray}
|\Phi_{1}\rangle_{a_{1}b_{1}d_{1}d_{2}}&=&\frac{1}{\sqrt{2}}(|N,0,1,0\rangle+|0,N,1,0\rangle)_{a_{1}b_{1}d_{1}d_{2}}
+\frac{1}{\sqrt{2}}(|N,0,0,1\rangle-|0,N,0,1\rangle)_{a_{1}b_{1}d_{1}d_{2}}.\label{BS}
\end{eqnarray}

 If the detector D$_{1}$ fires, Eq. ({\ref{BS}}) will collapse to
\begin{eqnarray}
|\phi_{1}\rangle_{a_{1}b_{1}}&=&\frac{1}{\sqrt{2}}(|N,0\rangle_{a_{1}b_{1}}+|0,N\rangle_{a_{1}b_{1}}),\label{max}
\end{eqnarray}
while if the detector D$_{2}$ fires, Eq. ({\ref{BS}}) will collapse to
\begin{eqnarray}
|\phi'_{1}\rangle_{a_{1}b_{1}}&=&\frac{1}{\sqrt{2}}(|N,0\rangle_{a_{1}b_{1}}-|0,N\rangle_{a_{1}b_{1}}).\label{max1}
\end{eqnarray}

Both Eq. ({\ref{max}}) and Eq. ({\ref{max1}}) are the maximally entangled N-photon NOON states and there is only a phase difference between them. Eq. (\ref{max1}) can be converted to Eq. (\ref{max}) by the phase flip operation with the help of a half-wave plate. So far, our first concentration protocol is completed. In the protocol, with the help of the weak cross-Kerr nonlinearity and the BS, by selecting the items which make the coherent state pick up the phase shift $\pm\theta$, Alice and Bob can distill the maximally entangled N-photon NOON state with the success probability P=$2|\alpha\beta|^{2}$.

Moreover, we will prove that the ECP can be reused to further concentrate the discarded items. The discarded items which make the coherent state pick up no phase shift can be written as
\begin{eqnarray}
|\Phi_{2}\rangle_{a_{1}b_{1}a_{2}b_{2}}&=&\alpha^{2}|N,0,1,0\rangle_{a_{1}b_{1}a_{2}b_{2}}+\beta^{2}|0,N,0,1\rangle_{a_{1}b_{1}a_{2}b_{2}}.
\end{eqnarray}
Alice and Bob also make the photons in the spatial modes a$_{2}$ and b$_{2}$ pass through the BS, and they can finally get
\begin{eqnarray}
|\Phi_{2}\rangle_{a_{1}b_{1}d_{1}d_{2}}&=&(\alpha^{2}|N,0,1,0\rangle+\beta^{2}|0,N,1,0\rangle)_{a_{1}b_{1}d_{1}d_{2}}+(\alpha^{2}|N,0,0,1\rangle-\beta^{2}|0,N,0,1\rangle)_{a_{1}b_{1}d_{1}d_{2}}.\label{BS2}
\end{eqnarray}
It is obvious if the detector D$_{1}$ fires, Eq. ({\ref{BS2}}) will collapse to
\begin{eqnarray}
|\phi_{2}\rangle_{a_{1}b_{1}}&=&\alpha^{2}|N,0\rangle_{a_{1}b_{1}}+\beta^{2}|0,N\rangle_{a_{1}b_{1}},\label{new}
\end{eqnarray}
while if the detector D$_{2}$ fires, Eq. ({\ref{BS2}}) will collapse to
\begin{eqnarray}
|\phi'_{2}\rangle_{a_{1}b_{1}}&=&\alpha^{2}|N,0\rangle_{a_{1}b_{1}}-\beta^{2}|0,N\rangle_{a_{1}b_{1}}.\label{new1}
\end{eqnarray}

Similar with Eq. ({\ref{max}}) and Eq. ({\ref{max1}}), Eq. ({\ref{new1}}) can be easily converted to Eq. ({\ref{new}}) by passing it through a half-wave plate. We can find that Eq. ({\ref{new}}) has the similar form with Eq. ({\ref{oringin}}), that is to say, Eq. ({\ref{new}}) is a new less-entangled N-photon NOON state and can be reconcentrated for the next round. In the second concentration round, the single photon source S$_{2}$ emits an auxiliary photon with the form
\begin{eqnarray}
|\phi'\rangle_{a_{2}b_{2}}&=&\alpha^{2}|1,0\rangle_{a_{2}b_{2}}+\beta^{2}|0,1\rangle_{a_{2}b_{2}}.
\end{eqnarray}
Following the steps in the first concentration round, Bob makes the photons in the spatial modes b$_{1}$ and b$_{2}$ pass through the QND, and the whole N+1 photon system combined with the coherent state can be described as
 \begin{eqnarray}
|\Phi'\rangle_{a_{1}b_{1}a_{2}b_{2}}\otimes|\alpha\rangle&=& |\phi_{2}\rangle_{a_{1}b_{1}}\otimes |\phi'\rangle_{a_{2}b_{2}}\otimes|\alpha\rangle\nonumber\\
&\rightarrow&\alpha^{4}|N,0,1,0\rangle_{a_{1}b_{1}a_{2}b_{2}}|\alpha\rangle+\alpha^{2}\beta^{2}|N,0,0,1\rangle_{a_{1}b_{1}a_{2}b_{2}}|\alpha e^{i\theta}\rangle\nonumber\\
&+&\alpha^{2}\beta^{2}|0,N,1,0\rangle_{a_{1}b_{1}a_{2}b_{2}}|\alpha e^{-i\theta}\rangle+\beta^{4}|0,N,0,1\rangle_{a_{1}b_{1}a_{2}b_{2}}|\alpha\rangle.\label{whole1}
\end{eqnarray}

Then Bob still selects the items which make the coherent state pick up the phase shift $\pm\theta$, and the Eq. (\ref{whole1}) can collapse to Eq. (\ref{pick}). For obtaining the maximally entangled N-photon NOON state, Alice and Bob also make the photons in the spatial mode a$_{2}$ and b$_{2}$ pass through the BS. After the BS, Eq. (\ref{pick}) can evolve to Eq. ({\ref{max}}) and Eq. (\ref{max1}), which are the maximally entangled N-photon NOON states. Therefore, in the second concentration round, Eq. (\ref{new}) can be successfully converted to the maximally entangled N-photon NOON state with the probability P$_{2}=\frac{2|\alpha\beta|^{4}}{|\alpha|^{4}+|\beta|^{4}}$, where the subscript '2' means in the second concentration round.

Similarly, by making the discarded items in the second concentration round pass through the BS, it can ultimately evolve to
\begin{eqnarray}
|\phi_{3}\rangle_{a_{1}b_{1}}&=&\alpha^{4}|N,0\rangle_{a_{1}b_{1}}+\beta^{4}|0,N\rangle_{a_{1}b_{1}},\label{new2}
\end{eqnarray}
which is a new less-entangled state and can be reconcentrated in the third round. Therefore, in this way,  our ECP can be used repeatedly to further concentrate less-entangled NOON state.

\section{The second ECP of the N-photons NOON state}
\begin{figure}[!h]
\begin{center}
\includegraphics[width=8cm,angle=0]{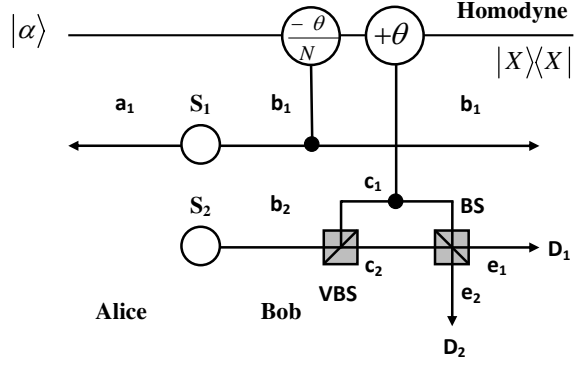}
\caption{A schematic diagram of our second ECP. The single photon source S$_{2}$ emits an auxiliary photon and only sends it to Bob. By passing it through the VBS, a single-photon entangled state can be created. Bob makes the photons in the mode b$_{1}$ and c$_{1}$ pass through two cross-Kerr nonlinearities, and selects the items which make the coherent state pick up the phase shift of $\pm\theta$. Then ,Bob makes the photons in the modes c$_{1}$ and c$_{2}$ pass through the BS and takes the Bell measurement. If a suitable VBS with the transmission of t$=\alpha^{2}$ can be provided, they can ultimately obtain the maximally entangled NOON state. Moreover, by choosing suitable VBS with the transmission of t$_{K}= \frac{|\alpha|^{2^{K}}}{|\alpha|^{2^{K}}+|\beta|^{2^{K}}}$, the ECP can be used repeatedly to get a high success probability.}
\end{center}
\end{figure}

The schematic diagram of our second ECP is shown in Fig. 2. We also suppose that Alice and Bob share a less-entangled N-photon NOON state $|\phi\rangle_{a_{1}b_{1}}$  with the form of Eq. (\ref{oringin}), which is generated by S$_{1}$. Here, single-photon source S$_{2}$ emits an auxiliary  photon and only sends it to Bob in the spatial mode b$_{2}$. Bob makes this auxiliary photon pass through a variable beam splitter (VBS) with the transmission of t, which can create an single-photon entangled state between the spatial modes c$_{1}$ and c$_{2}$ with the form
\begin{eqnarray}
|\psi\rangle_{c_{1}c_{2}}&=&\sqrt{1-t}|1,0\rangle_{c_{1}c_{2}}+\sqrt{t}|0,1\rangle_{c_{1}c_{2}}.\label{single}
 \end{eqnarray}

Then the whole N+1 photons system can be descried as
\begin{eqnarray}
|\psi\rangle_{a_{1}b_{1}c_{1}c_{2}}&=&|\phi\rangle_{a_{1}b_{1}}\otimes|\psi\rangle_{c_{1}c_{2}}.\nonumber\\
&=&\alpha\sqrt{1-t}|N,0,1,0\rangle_{a_{1}b_{1}c_{1}c_{2}}+\alpha\sqrt{t}|N,0,0,1\rangle_{a_{1}b_{1}c_{1}c_{2}}\nonumber\\
&+&\beta\sqrt{1-t}|0,N,1,0\rangle_{a_{1}b_{1}c_{1}c_{2}}+\beta\sqrt{t}|0,N,0,1\rangle_{a_{1}b_{1}c_{1}c_{2}}.\label{whole2}
\end{eqnarray}

Bob makes the photons in the spatial mode b$_{1}$ and c$_{1}$ pass through the QND, and the whole N+1 photon system combined with the coherent state can evolve to
\begin{eqnarray}
|\Psi\rangle_{a_{1}b_{1}c_{1}c_{2}}&=&|\psi\rangle_{a_{1}b_{1}c_{1}c_{2}}\otimes|\alpha\rangle\nonumber\\
&\rightarrow&\alpha\sqrt{1-t}|N,0,1,0\rangle_{a_{1}b_{1}c_{1}c_{2}}|\alpha e^{i\theta}\rangle+\alpha\sqrt{t}|N,0,0,1\rangle_{a_{1}b_{1}c_{1}c_{2}}|\alpha\rangle\nonumber\\
&+&\beta\sqrt{1-t}|0,N,1,0\rangle_{a_{1}b_{1}c_{1}c_{2}}|\alpha\rangle+\beta\sqrt{t}|0,N,0,1\rangle_{a_{1}b_{1}c_{1}c_{2}}|\alpha e^{-i\theta}\rangle.\label{whole3}
\end{eqnarray}

It can be seen that the items $\alpha\sqrt{1-t}|N,0,1,0\rangle_{a_{1}b_{1}c_{1}c_{2}}$ and $\beta\sqrt{t}|0,N,0,1\rangle_{a_{1}b_{1}c_{1}c_{2}}$ can make the coherent state pick up the phase shift of $\pm\theta$, while both the items $\alpha\sqrt{t}|N,0,0,1\rangle_{a_{1}b_{1}c_{1}c_{2}}$ and $\beta\sqrt{1-t}|0,N,1,0\rangle_{a_{1}b_{1}c_{1}c_{2}}$ make it pick up no phase shift. Bob selects the items corresponding to the phase shift $\pm\theta$ and discards other items. Then Eq. (\ref{whole2}) can collapse to
\begin{eqnarray}
|\Psi_{1}\rangle_{a_{1}b_{1}c_{1}c_{2}}&=&\alpha\sqrt{1-t}|N,0,1,0\rangle_{a_{1}b_{1}c_{1}c_{2}}\nonumber+\beta\sqrt{t}|0,N,0,1\rangle_{a_{1}b_{1}c_{1}c_{2}},\label{pick3}
\end{eqnarray}
with the success probability of $2|\alpha\beta|^{2}$.

Then, Bob makes the photons in the spatial modes c$_{1}$ and c$_{2}$ pass through the 50:50 BS, which makes
\begin{eqnarray}
\hat{c}_{1}^{\dagger}|0\rangle=\frac{1}{\sqrt{2}}(\hat{e}_{1}^{\dagger}|0\rangle+\hat{e}_{2}^{\dagger}|0\rangle)\nonumber\\ \hat{c}_{2}^{\dagger}|0\rangle=\frac{1}{\sqrt{2}}(\hat{e}_{1}^{\dagger}|0\rangle-\hat{e}_{2}^{\dagger}|0\rangle).
\end{eqnarray}
After the BS, Eq. ({\ref{pick3}}) can evolve to
\begin{eqnarray}
|\Psi_{1}\rangle_{a_{1}b_{1}e_{1}e_{2}}&=&\frac{\alpha\sqrt{1-t}}{\sqrt{2}}|N,0,1,0\rangle_{a_{1}b_{1}e_{1}e_{2}}+\frac{\beta\sqrt{t}}{\sqrt{2}}|0,N,1,0\rangle_{a_{1}b_{1}e_{1}e_{2}}\nonumber\\
&+&\frac{\alpha\sqrt{1-t}}{\sqrt{2}}|N,0,0,1\rangle_{a_{1}b_{1}e_{1}e_{2}}-\frac{\beta\sqrt{t}}{\sqrt{2}}|0,N,0,1\rangle_{a_{1}b_{1}e_{1}e_{2}}.\label{BS4}
\end{eqnarray}
It is obvious that if the detector D$_{1}$ fires, Eq. (\ref{BS4}) will collapse to
\begin{eqnarray}
|\psi_{1}\rangle_{a_{1}b_{1}}&=&\alpha\sqrt{1-t}|N,0\rangle_{a_{1}b_{1}}+\beta\sqrt{t}|0,N\rangle_{a_{1}b_{1}},\label{max2}
 \end{eqnarray}
while if the detector D$_{2}$ fires, Eq. (\ref{BS4}) will evolve to
\begin{eqnarray}
|\psi_{2}\rangle_{a_{1}b_{1}}&=&\alpha\sqrt{1-t}|N,0\rangle_{a_{1}b_{1}}-\beta\sqrt{t}|0,N\rangle_{a_{1}b_{1}}.\label{max3}
 \end{eqnarray}
If Bob gets Eq. (\ref{max3}), he can convert it to Eq. (\ref{max2}) by passing it through the half-wave plate. According to Eq. (\ref{max2}), if we can find a suitable VBS with the transmission $t=\alpha^{2}$, Eq. (\ref{max2}) can evolve to Eq. (\ref{max}), which is the maximally entangled NOON state. So far, our second ECP is completed. With the help of the weak cross-Kerr nonlinearity and the VBS, we can ultimately distill the maximally entangled N-photon NOON state with the probability of P=$2|\alpha\beta|^{2}$.

Moreover, we can also prove that the second ECP can be used repeatedly. When $t=\alpha^{2}$, the discarded items in the first concentration round can be written as
\begin{eqnarray}
|\Psi_{2}\rangle_{a_{1}b_{1}c_{1}c_{2}}&=&\alpha^{2}|N,0,0,1\rangle_{a_{1}b_{1}c_{1}c_{2}}+\beta^{2}|0,N,1,0\rangle_{a_{1}b_{1}c_{1}c_{2}}.\label{discard}
\end{eqnarray}
Bob still makes the photons in the spatial modes c$_{1}$ and c$_{2}$ pass through the BS, and Eq. (\ref{discard}) can evolve to
\begin{eqnarray}
 |\Psi_{2}\rangle_{a_{1}b_{1}e_{1}e_{2}}&=&\frac{\alpha^{2}}{\sqrt{2}}|N,0,1,0\rangle_{a_{1}b_{1}e_{1}e_{2}}+\frac{\beta^{2}}{\sqrt{2}}|0,N,1,0\rangle_{a_{1}b_{1}e_{1}e_{2}}\nonumber\\
&+&\frac{\alpha^{2}}{\sqrt{2}}|N,0,0,1\rangle_{a_{1}b_{1}e_{1}e_{2}}-\frac{\beta^{2}}{\sqrt{2}}|0,N,0,1\rangle_{a_{1}b_{1}e_{1}e_{2}}.\label{discard1}
\end{eqnarray}
Therefore, if the detector D$_{1}$ fires, Eq. (\ref{discard1}) will collapse to Eq. (\ref{new}), while if the detector D$_{2}$ fires, Eq. (\ref{discard1}) will collapse to Eq. (\ref{new1}). Eq. (\ref{new1}) can be converted to Eq. (\ref{new}) by the phase flip operation. According to the description in Sec. II, Eq. (\ref{new}) is the new less entangled N-photon NOON states and can be reconcentrated in the next round.

In the second concentration round, the single photon source S$_{2}$ emits another auxiliary photon and sends it to Bob. By passing it through the VBS, a new single-photon  entangled  state with the form of Eq. (\ref{single}) can be created. Bob still makes the photons in the spatial modes b$_{1}$ and c$_{1}$ pass through the QND. Then the whole N+1 photons system combined with the coherent state can be written as
 \begin{eqnarray}
|\psi'\rangle_{a_{1}b_{1}c_{1}c_{2}}&=&|\phi_{2}\rangle_{a_{1}b_{1}}\otimes|\psi\rangle_{c_{1}c_{2}}\otimes|\alpha\rangle\nonumber\\
&\rightarrow&\alpha^{2}\sqrt{1-t}|N,0,1,0\rangle_{a_{1}b_{1}c_{1}c_{2}}|\alpha e^{i\theta}\rangle+\alpha^{2}\sqrt{t}|N,0,0,1\rangle_{a_{1}b_{1}c_{1}c_{2}}|\alpha\rangle\nonumber\\
&+&\beta^{2}\sqrt{1-t}|0,N,1,0\rangle_{a_{1}b_{1}c_{1}c_{2}}|\alpha\rangle+\beta^{2}\sqrt{t}|0,N,0,1\rangle_{a_{1}b_{1}c_{1}c_{2}}|\alpha e^{-i\theta}\rangle.\label{whole4}
\end{eqnarray}
Bob still selects the items which make the coherent state pick up the phase shift $\pm\theta$, then Eq. (\ref{whole4}) can evolve to
 \begin{eqnarray}
|\Psi_{1}'\rangle_{a_{1}b_{1}c_{1}c_{2}}&=&\alpha^{2}\sqrt{1-t}|N,0,1,0\rangle_{a_{1}b_{1}c_{1}c_{2}}+\beta^{2}\sqrt{t}|0,N,0,1\rangle_{a_{1}b_{1}c_{1}c_{2}},\label{pick4}
\end{eqnarray}
with the success probability of $\frac{2|\alpha\beta|^{4}}{|\alpha|^{4}+|\beta|^{4}}$.

Next, Bob makes the photons in the spatial modes c$_{1}$ and c$_{2}$ pass through the BS, and the Eq. (\ref{pick4}) can finally evolve to
\begin{eqnarray}
|\psi_{1}'\rangle_{a_{1}b_{1}}&=&\alpha^{2}\sqrt{1-t}|N,0\rangle_{a_{1}b_{1}}+\beta^{2}\sqrt{t}|0,N\rangle_{a_{1}b_{1}}.\label{max4}
 \end{eqnarray}
\begin{figure}[!h]
\begin{center}
\includegraphics[width=8cm,angle=0]{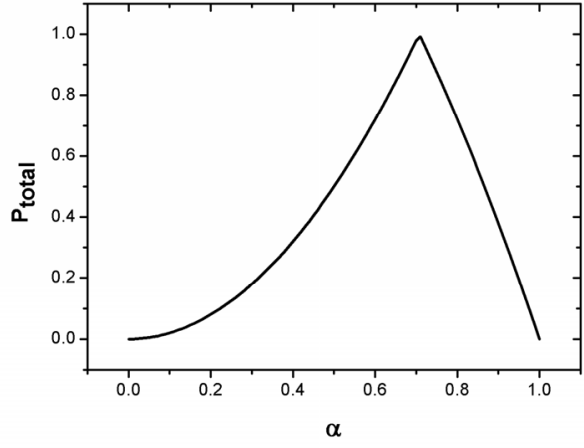}
\caption{The success probability ($P_{total}$) for obtaining a maximally entangled N-photon NOON state after the concentration protocol being operated for K times. For numerical simulation, we choose $K=10$. It can be seen that the value of $P_{total}$ largely depends on the initial coefficient $\alpha$. When $\alpha=\frac{1}{\sqrt{2}}$, $P_{total}$ reaches the maximum as 1.}
\end{center}
\end{figure}
If we can find a suitable VBS with the transmission $t_{2}=\frac{|\alpha|^{4}}{|\alpha|^{4}+|\beta|^{4}}$, where the subscript '2' means in the second concentration round, Eq. (\ref{max4}) can evolve to Eq. (\ref{max}). That is to say, Eq. (\ref{new}) can be ultimately recovered to the maximally entangled NOON state with the probability P$_{2}=\frac{2|\alpha\beta|^{4}}{|\alpha|^{4}+|\beta|^{4}}$. Similarly, by passing the discarded items in the second concentration round through the BS, the discarded items can also be evolve to Eq. (\ref{new2}), which can be reconcentrated in the next round. Therefore, by choosing the suitable VBS with the transmission  t$_{K}= \frac{|\alpha|^{2^{K}}}{|\alpha|^{2^{K}}+|\beta|^{2^{K}}}$, where the subscript 'K' means the iteration time, the second ECP can be used repeatedly to distill the maximally entangled NOON state.

Finally, we would like to calculate the total success probability of these two ECPs. Both the two ECPs have the same success probability. In each concentration round, the success probability can be shown as
\begin{eqnarray}
P_{1}&=&2|\alpha\beta|^{2}\nonumber\\
P_{2}&=&\frac{2|\alpha\beta|^{4}}{|\alpha|^{4}+|\beta|^{4}}\nonumber\\
P_{3}&=&\frac{2|\alpha\beta|^{8}}{(|\alpha|^{4}+|\beta|^{4})(|\alpha|^{8}+|\beta|^{8})}\nonumber\\
P_{4}&=&\frac{2|\alpha\beta|^{16}}{(|\alpha|^{4}+|\beta|^{4})(|\alpha|^{8}+|\beta|^{8})(|\alpha|^{16}+|\beta|^{16})}\nonumber\\
&\cdots\cdots&\nonumber\\
P_{K}&=&\frac{2|\alpha\beta|^{2^{K}}}{(|\alpha|^{4}+|\beta|^{4})(|\alpha|^{8}+|\beta|^{8})\cdots(|\alpha|^{2^{K}}+|\beta|^{2^{K}})}.\\\label{probability}
\end{eqnarray}

In theory, the ECPs can be reused indefinitely, the total success probability for distilling the maximally entangled N-photon NOON state from the less-entangled NOON state equals the sum of the probability in each concentration round.
\begin{eqnarray}
P_{total}=P_{1}+P_{2}+\cdots P_{K}=\sum\limits_{K=1}^{\infty} P_{K}
\end{eqnarray}

It can be found that if the initial entangled state is the maximally entangled NOON state, where $\alpha=\beta=\frac{1}{\sqrt{2}}$, the probability $P_{total}=\lim(\frac{1}{2}+\frac{1}{4}+\frac{1}{8}+\cdots\frac{1}{2^{K}}+\cdots)=1$, while if $\alpha\neq\beta$, the $P_{total}<1$. Here, we choose $K=10$ as a proper approximation and calculate the value of $P_{total}$ in different initial entanglement coefficient. Fig. 3 shows the value of $P_{total}$ as a function of the entanglement coefficient $\alpha$. It can be found that $P_{total}$ largely depends on the initial entanglement state. The higher initial entanglement can lead to the greater $P_{total}$.

\section{Discussion and summary}

In the paper, we presented two efficient ECPs for distilling the maximally entangled N-photon NOON state from arbitrary less-entangled NOON state. In our ECPs, only a pair of less-entangled NOON state and an  auxiliary single photon are required to complete the task.
It is interesting to compare these two ECPs with other ones. In most conventional ECPs, they require two similar copies of less-entangled states to
obtain one pair of maximally entangled state. Our two ECPs only require one pair of less-entangled state, and can reach the same high success probability as the convention ECPs. With the help of cross-Kerr nonlinearity, after performing these ECPs, both the maximally entangled NOON state and the less-entangled NOON state can be remained for various applications. During these ECPs, they only require the weak cross-Kerr nonlinearity, which can greatly reduce the experimental difficulties. As pointed by Refs. \cite{gate,gate1}, in a practical operation, we should require $\alpha\theta>1$  to make the phase shift reach an observed value. So
in the condition of weak cross-Kerr nonlinearity ($\theta$ is small), the requirement can be satisfied by large amplitude of the coherent state.
Certainly, both ECPs require to know the initial coefficients $\alpha$ and $\beta$ in advance. Meanwhile, they also should make the coherent state $|\alpha\rangle$ pick up different phase shifts, with one is $-\frac{\theta}{N}$  and the other is $\theta$.
According to the relationship of $\theta=\chi t$, it can be realized by selecting the different Kerr materials or controlling the coupling time $t$.

In the first ECP, the auxiliary photon should be shared by Alice and Bob in distant locations.  Moreover, after passing through the QND, the auxiliary photon should be sent back and will be transmitted in a second time. However, as pointed out by Ref. \cite{memory}, the photon loss is unavoidable during the transmission process due to the noisy environment. It will greatly limit this ECP in the practical application in quantum information processing. Fortunately, in the second ECP, the auxiliary photon entanglement can be created locally and only possessed by Bob, which makes this ECP more powerful than the first one. Moreover, only Bob needs to operate this concentrate step, which leads this ECP more simple to be realized.
In the second ECP, the VBS is the key element for creating the local single-photon entanglement. According to the initial coefficients of the less-entangled
 NOON state, they should adjust the different transmission t to satisfy t$_{K}= \frac{|\alpha|^{2^{K}}}{|\alpha|^{2^{K}}+|\beta|^{2^{K}}}$ in the K$th$ concentration round. Certainly, in the first ECP, the single-photon entanglement created by S$_{2}$ can also be reviewed as the single-photon source plus the VBS. The VBS is a common optical element in current technology.
 Recently, using the VBS, Osorio \emph{et al.} reported their experimental results about the heralded photon amplification for quantum communication \cite{amplification}. They adjusted the transmission of VBS to increase the probability of the single photon $|1\rangle$ from a mixed state $\eta_{t}|1\rangle\langle1|+(1-\eta_{t})|0\rangle\langle0|$.
   In their protocol, they adjust the splitting ratio of VBS from 50:50 to 90:10 to increase
   the visibility from 46.7 $\pm$ 3.1\% to 96.3 $\pm$ 3.8\%.

In summary, we put forward two efficient ECPs for distilling the maximally entangled N-photon NOON state from arbitrary less-entangled NOON state. In both two ECPs, we only require a pair of less-entangled NOON state and an auxiliary single photon to complete the task.  These two ECPs have two  advantages: First, they do not require the large cross-Kerr nonlinearity, and the weak cross-Kerr nonlinearity can achieve the task. Second, with the help of the weak cross-Kerr nonlinearities, both two ECPs can be used repeatedly and have the same high success probability. Moreover, the second ECP is more optimal than the first one for two main reasons. First, all the operations in the second ECP are local, while those in the first ECP are nonlocal. Second, in the second protocol, benefit from the VBS, the requirement for the single-photon source is much lower. The two advantages can simply the operation and make the second ECP can be realized much easier under present experimental conditions. Therefore, our two ECPs, especially the second ECP may be useful and convenient in the current quantum information processing.

\section*{ACKNOWLEDGEMENTS}
 This work is supported by the National Natural Science Foundation of
China under Grant No. 11104159,  Open Research
Fund Program of the State Key Laboratory of
Low-Dimensional Quantum Physics Scientific, Tsinghua University,
Open Research Fund Program of National Laboratory of Solid State Microstructures  under Grant No. M25020 and M25022, Nanjing University,
Scientific Research
Foundation of Nanjing University of Posts and Telecommunications
under Grant No. NY211008,  and the Project
Funded by the Priority Academic Program Development of Jiangsu
Higher Education Institutions.

{}

\end{document}